\documentclass[graybox]{svmult}

\usepackage{type1cm}        % activate if the above 3 fonts are
\usepackage{makeidx}         % allows index generation
\usepackage{graphicx}        % standard LaTeX graphics tool
\usepackage{multicol}        % used for the two-column index
\usepackage[bottom]{footmisc}% places footnotes at page bottom

\usepackage{newtxtext}       % 
\usepackage{newtxmath}       % selects Times Roman as basic font

\makeindex             % used for the subject index
\begin{document}

\title*{Classification of Stress via Ambulatory ECG and GSR Data}
\author{Zachary Dair, Muhammad Saad, Urja Pawar, Samantha Dockray, Ruairi O’Reilly}
\institute{Zachary Dair \at Munster Technological University, Ireland, \email{zachary.dair@mycit.ie}
\and Muhammad Saad \at Munster Technological University, Ireland, \email{muhammad.saad@mycit.ie}
\and Urja Pawar \at Munster Technological University, Ireland, \email{urja.pawar@mycit.ie}
\and Samantha Dockray \at University College Cork, Ireland, \email{s.dockray@ucc.ie}
\and Ruairi O’Reilly \at Munster Technological University, Ireland, \email{ruairi.oreilly@mtu.ie}}
%
% Use the package “url.sty” to avoid
\maketitle

\abstract*{In healthcare, detecting stress and enabling individuals to monitor their mental health and well-being is challenging. Advancements in wearable technology now enable continuous physiological data collection. This data can provide insights into mental health and behavioural states through psychophysiological analysis. However, automated analysis is required to provide timely results due to the quantity of data collected. Machine learning has shown efficacy in providing an automated classification of physiological data for health applications in controlled laboratory environments. Ambulatory uncontrolled environments, however, provide additional challenges requiring further modelling to overcome. This work empirically assesses several approaches utilising machine learning classifiers to detect stress using physiological data recorded in an ambulatory setting with self-reported stress annotations. A subset of the training portion of the SMILE dataset enables the evaluation of approaches before submission. The optimal stress detection approach achieves 90.77\% classification accuracy, 91.24 F1-Score, 90.42 Sensitivity and 91.08 Specificity, utilising an ExtraTrees classifier and feature imputation methods. Meanwhile, accuracy on the challenge data is much lower at 59.23\% (submission \#54 from BEaTS-MTU, username ZacDair). The cause of the performance disparity is explored in this work.}

\abstract{In healthcare, detecting stress and enabling individuals to monitor their mental health and well-being is challenging. Advancements in wearable technology now enable continuous physiological data collection. This data can provide insights into mental health and behavioural states through psychophysiological analysis. However, automated analysis is required to provide timely results due to the quantity of data collected. Machine learning has shown efficacy in providing an automated classification of physiological data for health applications in controlled laboratory environments. Ambulatory uncontrolled environments, however, provide additional challenges requiring further modelling to overcome. This work empirically assesses several approaches utilising machine learning classifiers to detect stress using physiological data recorded in an ambulatory setting with self-reported stress annotations. A subset of the training portion of the SMILE dataset enables the evaluation of approaches before submission. The optimal stress detection approach achieves 90.77\% classification accuracy, 91.24 F1-Score, 90.42 Sensitivity and 91.08 Specificity, utilising an ExtraTrees classifier and feature imputation methods. Meanwhile, accuracy on the challenge data is much lower at 59.23\% (submission \#54 from BEaTS-MTU, username ZacDair). The cause of the performance disparity is explored in this work.}

\section{Introduction}
\label{sec:1}
Stress is a psychophysiological reaction in response to internal or external stressors. In psychological models of behaviour, stress can be represented by considerations of arousal, as the level of activation, and valence, as the negative or positive dimension of the emotional state\cite{johnson_1990_stress}. A moderate stress level can support people to develop resilience and respond to the demands of the situation. However, high acute or prolonged stress can present a risk to mental health, and is linked to physical illness. While high stress over a prolonged period can lead to harmful effects such as depression, mental disorders, reduced job productivity, and enhanced risk of various somatic and mental illnesses \cite{yu2022semi,garg2021stress}.

These conditions can degrade physical and mental health if not treated, however effective and timely treatment may rely on detection via accurate monitoring. Providing individuals with accessible, continuous stress monitoring delivers greater awareness and insights into daily stress enabling self-moderation, distancing from stressors, and ultimately leading to reduced levels of daily stress.

Biomedical healthcare devices such as wearable sensors provide individuals with physiological monitoring capabilities. These devices contain multiple sensors capable of recording various physiological signals, such as electrocardiograms (ECG) and galvanic skin response/electrodermal activity (GSR/EDA).

ECG measures the heart’s electrical activity, commonly utilised in a medical setting for arrhythmia detection \cite{LUZ2016144}. Additionally, ECG signals have been used to indicate psychological states based on the relationship of the psychological state to the autonomic nervous system, as articulated in the Polyvagal Theory \cite{Porges2009}, which is responsible for involuntary physiological responses due to psychological processes.

GSR measures the skin’s electrical activity variances due to fluctuating sweat levels that originate from the autonomic activation of sweat glands. GSR commonly represents an individual’s level of activation (Arousal), making it suitable for detecting stress and other high arousal psychological states\cite{bakker_stress_gsr}. These signals have demonstrated high efficacy for fitness tracking applications, physical health monitoring and are beginning to show potential for mental health monitoring \cite{Henriksen_fitness,Akane_stress_mental_health}. Such signals are often collected and analysed in widely available commercial wearable health monitors such as Samsung Galaxy Smartwatches, Fitbit, Apple Watch and research devices such as Empatica.

This work contributes a high-performing stress detection approach using ambulatory physiological data. This is achieved by (i) assessing the most suitable features indicative of stress for hourly and per-minute stress detection, by (ii) approximating missing data to reduce the impact on classification performance due to simulated sensor detachment, and by (iii) investigating disparity between training and testing data impeding the generalisability of a model.

\section{Challenges}
\label{sec:2}

The SMILE \cite{yang2022more} dataset highlights several commonly encountered challenges for conducting stress detection from wearable data gathered in an ambulatory setting. The most prominent challenge is the complexity of accurately detecting a fluctuation in psychological states based on physiological signals, which may fluctuate as a correlate of psychological states, including stress or other factors. This requires a granular analysis of physiological signals, in this case, ECG and GSR, to identify the baseline values for the individual and, subsequently, any fluctuations that could originate from involuntary reactions to stressors. Additionally, these fluctuations may originate from factors other than stress, such as movement artefacts, sensor error, or other psychological reactions requiring further modelling to avoid misclassification.

An additional challenge relates to the labelling of the data, specifically using one label per hour. Detecting psychological states from physiological signals commonly utilises shorter data windows to provide a granular analysis of the individual. With an hourly label of stress, it is unclear as to the actual duration of the stress reaction of the individual. As such, the classifier may train or predict features indicative of a neutral state that instead contain a stress label or inversely. Without knowing the ground truth label per minute, an approximation based on the hour can lead to potential miss-classifications.

Finally, missing data is a significant issue in the ambulatory assessment of individuals, often stemming from sensor detachment or movement artefacts. SMILE demonstrates this issue through the absence of certain features throughout the train and test data. Interestingly, both portions of the dataset include several instances (Train: 19 of 2070, Test: 2 of 986) with no data. Missing data is a significant aspect of this dataset. In the training set, a mere 17 of 2070 instances contain features for the entire hour, emphasising a requirement to use each modality and potential feature imputation to replace missing values.

\section{Methodology}
\begin{figure*}[h]
    \centering
    \includegraphics[width=\textwidth]{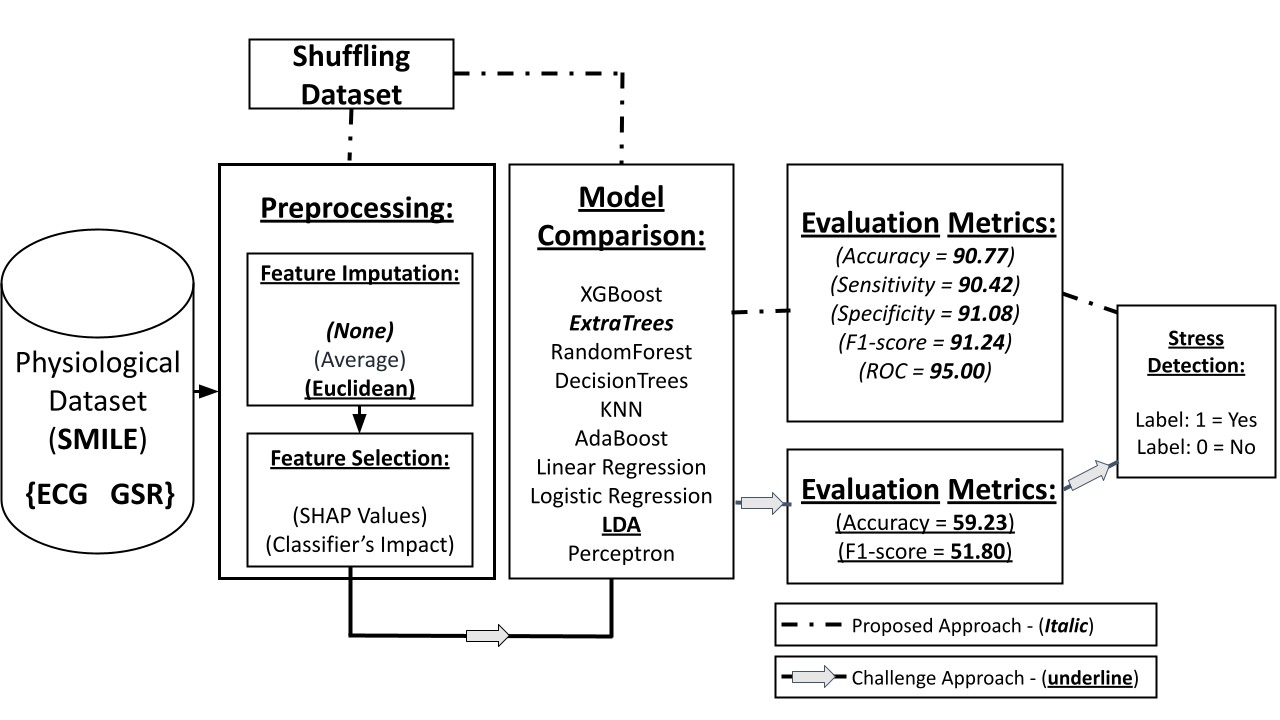}
    \caption{A detailed overview of the methodology for stress detection using physiological data.}
    \label{fig:flowdiagram}
\end{figure*}

The adopted methodology is depicted in Figure \ref{fig:flowdiagram}; two distinct workflows are denoted, the proposed approach and the challenge approach. In addition, the diagram demonstrates the components which contributed to the optimal approach for stress detection on SMILE data. 

\subsection{Dataset}
This study uses the SMILE dataset \cite{yang2022more}, which contains physiological data and self-reported stress annotations from 45 healthy participants (39 females and 6 males). Physiological data originates from two sensors - a wrist-worn device for GSR/EDA and Skin Temperature (ST) and a chest-worn device for ECG. Each dataset annotation is a self-reported stress level on a scale of 0 (not at all) to 6 (very) collected via a mobile phone application with random notifications prompting a report.

SMILE provides three categories of features, all extracted from 5-minute sliding windows with 4-minute overlapping segments:
Eight handcrafted ECG features are provided, consisting of a range of statistical and frequency measures, commonly known as heart rate variability measures.
Handcrafted GSR features are eight similar statistical and frequency measures derived from the GSR signal, and included in this category are four statistical measures of skin temperature.

There are two categories distinguished as deep features, and both originate from using an ECG signal in conjunction with a deep model, specifically a convolutional neural network and a transformer. These deep features are reconstructed portions of the original ECG signal for that period. 
Notably, the SMILE dataset provides specific training and testing portions of data. However, the testing set contains no labels, requiring the usage of an 80:20 split of training data to enable model evaluation. Subsequently, high-performing models are re-trained on the entire training set before classifying the test data.

\begin{table}[!t]
\caption{Percentage of data missing per hour and per minute in SMILE}
\label{tab:dataset_outline}
\begin{tabular}{p{2.7cm}p{2.6cm}p{2.6cm}p{2.7cm}}
\hline\noalign{\smallskip}
 Data Split & Feature & Minutes & Full Hours\\
\noalign{\smallskip}\svhline\noalign{\smallskip}
Train & ECG HC & 11.75\% & 10.96\% \\
Train & GSR HC & 12.59\% & 12.22\% \\
Train & Deep ECG & 20.67\% & 9.56\% \\
Train & All Features & 1.27\% & 0.91\% \\
\noalign{\smallskip}\svhline\noalign{\smallskip}
Test & ECG HC & 5.00\% & 4.15\% \\
Test & GSR HC & 6.79\% & 6.49\% \\
Test & Deep ECG & 24.72\% & 9.83\% \\
Test & All Features & 0.25\% & 0.20\% \\
\noalign{\smallskip}\hline\noalign{\smallskip}
\end{tabular}
\end{table}

\subsection{Feature Imputation}
In ambulatory physiological data recording, missing values due to sensor detachment is a common issue and causes significant issues for ML training and classification. Therefore, various methods of feature imputation are utilised to replace the missing data with appropriate values. A primitive method is to compute the average value of the missing feature across the entire dataset and substitute missing values with this average. However, this approach loses the inherent links between that feature and its label. A more complex method utilises Euclidean distance or other similarity measures on an existing feature to find a similar instance. Once a similar instance based on a single feature or combination of features is identified, the previously missing features are replaced with values found from the similar instance. This approach also has limitations, such as requiring a minimum of a single feature to be present to act as a comparator. 

\subsection{Feature Selection}
Feature selection is a pre-processing procedure to identify the most performant features in the data. For this work, an initial classification is conducted using the baseline classifiers and each feature individually to provide insights on per-feature performance for stress detection. However, as the dataset contains instances of missing data for all features, an ensemble of independent feature classifiers enables leveraging the individual potential of each feature. Furthermore, aggregating predictions and using a voting mechanism can compensate for missing data. Subsequently, each permutation of features is analysed to assess the potential for a high-performing combination of features. Finally, a feature vector containing all features is utilised to ensure that any latent patterns between features are captured and leveraged for stress detection. 

Another technique to perform feature selection is via feature importance. Feature importance techniques assign a score to participating features based on their impact on the model’s classification process and have been used in modelling many medical datasets \cite{pawar2020incorporating, pawar2021evaluating}. In this work, SHAP values\cite{lundberg2017unified} are used to analyse the importance of different features. SHAP values are based on the concept of Shapley, which is a game theory approach to identify the contribution of each player participating in a game towards achieving a goal \cite{lundberg2017unified}.

\subsection{Baseline Classifiers}
A battery of machine learning models of various architectures provides the initial baseline results of each new experimental procedure. Additionally, the variation in architectures generates insights surrounding the most appropriate classifier type for stress detection from ECG and GSR features.

\begin{itemize}
\item Linear Regression: A regression algorithm used to predict target value based on a set of independent features. This algorithm assumes a linear relationship between dependent and independent variables. Therefore, not suited for modelling non-linear relationships.
\item Logistic Regression: A classification algorithm used to predict binary outcomes from a set of independent features. It is based on a logarithmic link function that enables modelling non-linear associations in a linear way. While it does not assume a linear relationship between dependent and independent variables, it assumes a linear relationship between link function and independent variables that limits the effectivity of this algorithm in real-world non-linear data. 
\item K-Nearest Neighbours (KNN): Is used to predict the target class based on the existing samples in the training data by utilising different distance metrics (Euclidean, Manhattan, Hamming, Minkowski) in order to classify similar samples to a particular class. As the algorithm is based on finding similar samples, it does not perform well with too many features as it becomes harder to find similar samples as per a given distance metric. Also, it’s required to have homogeneous (same scale) features in the data.
\item  Linear Discriminant Analysis (LDA): A dimensionality reduction algorithm that can be used as a robust classification algorithm as it reduces high dimensional data to low dimensions that leads to an increase in classification performance by avoiding the curse of dimensionality. While normalised data is assumed for this algorithm to work well, there are ways to arrive at the same LDA features without normalised data. However, the algorithm constructs a linear decision boundary for classification, limiting its application to non-linear classification problems.   
\item Extra Trees: Extra trees is a tree ensemble algorithm that uses many trees to classify an input and decide classification based on majority voting. Each tree is a decision tree that learns to classify data upon conditioning on features. Each tree in the extra trees algorithm uses original training inputs of data with randomly assigned split conditions that enable a fast and effective classification process.
\item Random Forest: Random forest is also a tree ensemble algorithm that utilises many decision trees to classify an input instance. Random forest uses different replicas of the original training input with data replacement and optimum split conditions to train the trees. Therefore, random forests often perform better than other ML classifiers.
\item AdaBoost: Adaptive boosting is a boosting algorithm that combines a number of weak classifiers to generate one strong classifier. Initially, data is passed to the first model. Then, the incorrectly classified instances are assigned higher weights and passed to a second model in series to decrease their chance of being incorrectly classified. This is repeated with a limited number of models until the combination classifies with sufficient accuracy. However, this algorithm has reduced performance on noisy datasets.
\item XGBoost: A boosting technique that combines weak tree-based classifiers to create a strong classifier. Incorrectly classified samples from one decision tree are passed to another after increasing their weights (the penalty for misclassifying). This algorithm has built-in regularisation that makes it less prone to overfitting. 
\item Multi-Layer Perceptron (MLP): A classic neural-networks-based classifier that passes input to layers of operations to learn granular information about the data that enables it to identify complex non-linear decision surfaces for classification purposes. Classifiers based on neural networks perform well with large input data. However, these classifiers are computationally expensive to train and require considerable model-tuning to achieve high performance.
\end{itemize}

\subsection{Evaluation Criteria}
Several metrics, such as accuracy, precision, recall, sensitivity, specificity, F-measure, receiver operating characteristic curve (ROC), and area under the curve (AUC) scores, are used to assess the performance of ML classifiers for the stress detection tasks. Additional metrics are utilised as solely utilising classification accuracy lacks class-specific information of the approach evaluated \cite{sharma2021comprehensive}. Therefore, recall, sensitivity, specificity, and F1 scores are recommended to assess the number of positive and negative predictions made for the targeted classes. Furthermore, classifiers can be fine-tuned using the ROC and AUC metrics as they provide reasonable compensation between the true positive rate (TPR) and the false-positive rate (FPR). This work uses accuracy, F1-measure, sensitivity, specificity, and ROC metrics to evaluate the ML classifiers for detecting stress from the physiological dataset.

\subsection{Per Minute Classification}
Commonly used approaches for detecting stress or other psychological states use a reduced signal duration, ranging from seconds to several minutes. \cite{mukherjee_real-time_2022,tucker_catching_2021,Santos_gsr} This provides the distinct advantage of requiring less data, reducing computational complexity. Additionally, a smaller duration enables more frequent classifications of stress or other emotions, reducing lag and providing greater momentary insights into an individual. As such, the hourly instances of data provided in SMILE are converted into per-minute samples. However, as the annotation frequency of SMILE remains at one stress label per hour, it is unclear whether every minute of data within the hour truly reflects the associated label. Specifically, the per-hour label can add to classification confusion, as certain portions of data within an hour are likely to be neutral, with sporadic and limited instances of stress, due to the nature of stress and daily occurrences.

Once converted into per-minute instances of data, classification is conducted as normal. However, the original testing labels are provided per hour, indicating the aggregation method requirement. Several voting mechanisms were created to combine per-minute classifications into hourly accurately. The first utilises the mean classification per hour with a dynamic threshold enabling the weighting of predictions. The second is a highly sensitive approach, which monitors for any classification of stress within an hour, and, if detected, reports the entire hour as stress. The final method utilises an isolated LDA classifier using 60 classifications as features to return a single per-hour binary result indicating the presence or absence of stress.

\subsection{Covariate Shift}
The similarity of training and testing data was analysed using the baseline classifiers. First, each data instance is relabelled with a new label distinguishing each data portion as training (1) or testing (0) data. Subsequently, the newly labelled portions of data are combined into a single dataset, which is shuffled to ensure randomisation. Finally, the combined dataset is split into 80\% training and 20\% testing, and each of the classifiers aims to detect whether the data originates from training or testing splits. Using this method, high performance from the classifiers indicates substantial variance between training and testing data, typically called covariate shift. An issue arises when the distribution of subsets of data is varied. Often covariate shifts will lead to substantial negative impacts on testing and production classifications, requiring weighting, normalisation or combining the training and testing data to provide the classifier with a broader perspective of data when training. \cite{dharani_covariate}.

\section{Results and Discussion}

\begin{table}[!t]
\caption{Feature selection classification metrics hourly and per minute grouped by feature type}
\label{tab:feature_selection_table}
\begin{tabular}{p{3cm}p{2.4cm}p{2cm}p{2cm}p{2cm}}
\hline\noalign{\smallskip}
Features & Best Classifier & Duration & Accuracy & F1-Score \\
\noalign{\smallskip}\svhline\noalign{\smallskip}
All & RandForest & H & 0.65 & 0.65 \\
    All & Extra Trees Ens. & H & 0.60 & 0.71 \\
    All & XGBoost & M & 0.69 & 0.73 \\
    \textbf{ECG} & \textbf{ExtraTrees} & \textbf{H} &  \textbf{0.73} & \textbf{0.71} \\
    ECG & ExtraTrees & M & 0.69 & 0.68 \\
    ECG, GSR & XGBoost & H & 0.65 & 0.70 \\
    \textbf{ECG, GSR} & \textbf{LogReg} & \textbf{M} & \textbf{0.70} & \textbf{0.72} \\
    ECG, GSR, ECG\_T & ExtraTrees & H & 0.69 & 0.70\\
    ECG, GSR, ECG\_T & RandForest & M & 0.68 & 0.71 \\
    ECG\_C & LogReg & H &  0.59 & 0.61 \\
    ECG\_C & LogReg & M & 0.62 & 0.71 \\
    ECG\_T & XGBoost & H & 0.59 & 0.60 \\
    ECG\_T & LDA & M & 0.61 & 0.69 \\
    GSR & ExtraTrees & H & 0.63 & 0.69 \\
    GSR & LogReg & M & 0.65 & 0.74 \\
\noalign{\smallskip}\hline\noalign{\smallskip}
\end{tabular}
\end{table}

\begin{figure}[h]
    \centering
    \includegraphics[width=\textwidth]{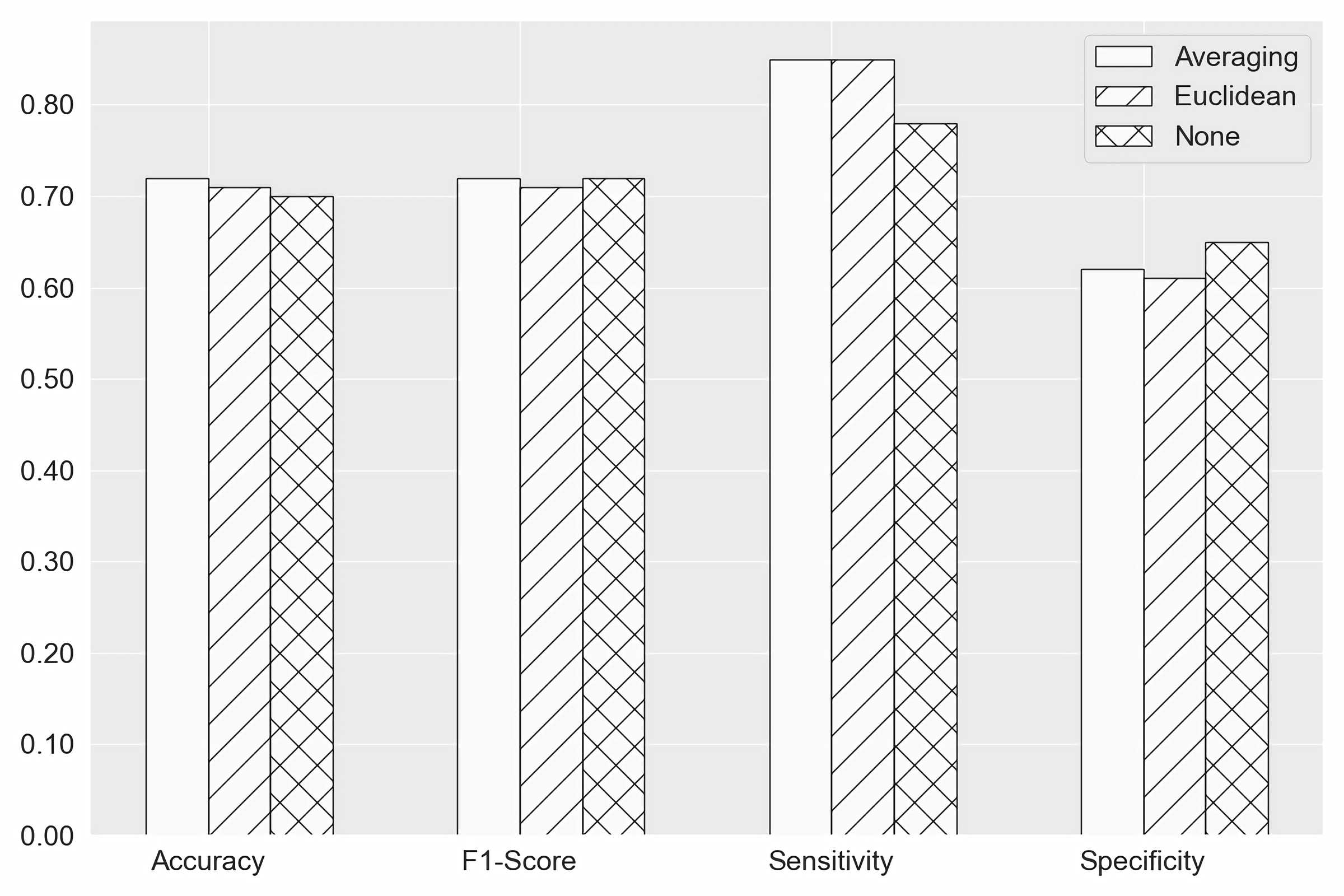}
    \caption{Impact of feature imputation methods on Logistic Regression}
    \label{fig:results_feature_imputation}
\end{figure}

\begin{figure}[h]
    \centering
    \includegraphics[width=\textwidth]{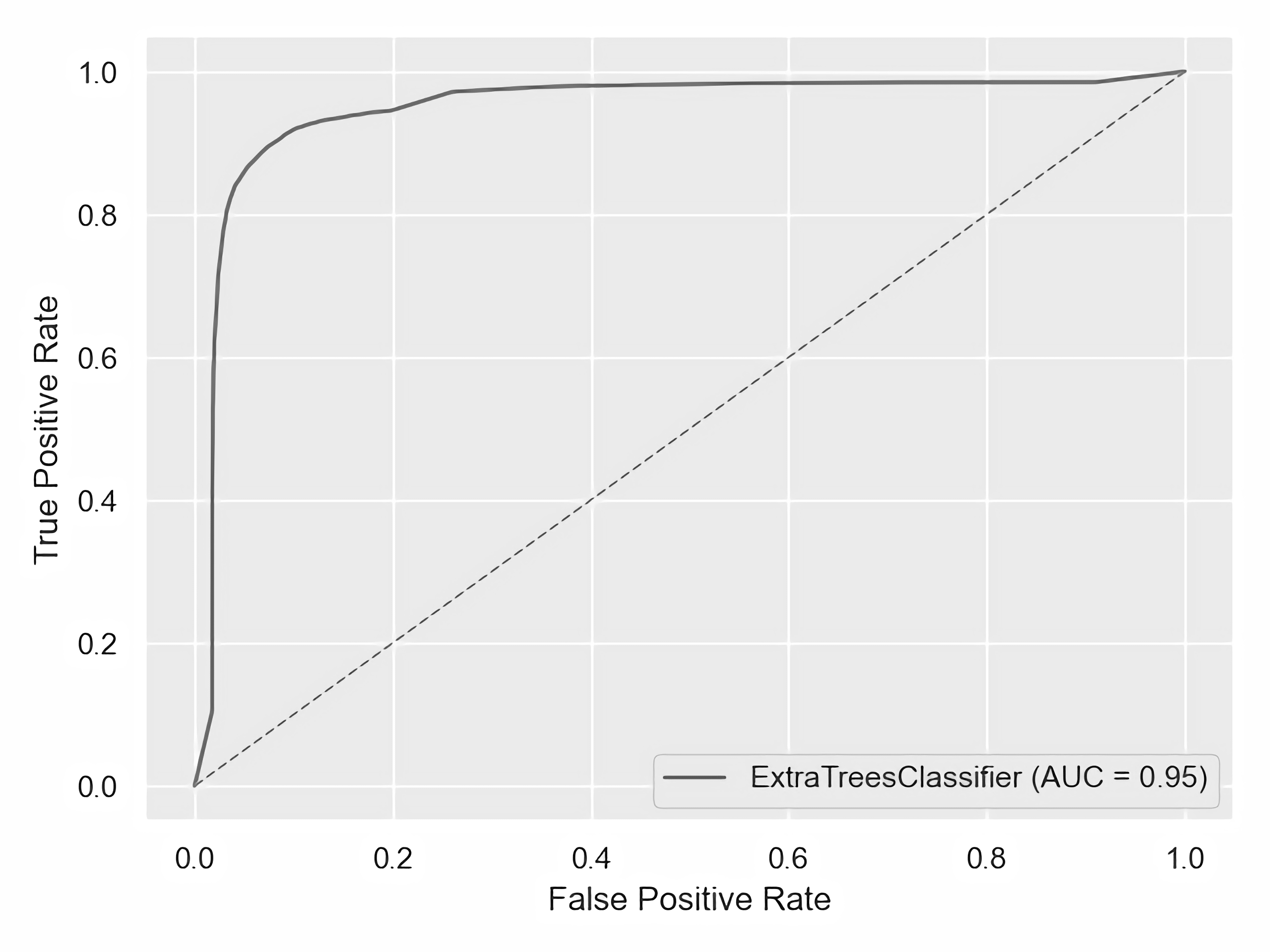}
    \caption{ROC curve for the highest performing approach}
    \label{fig:results_roc_optimal_approach}
\end{figure}

\begin{table}[!t]
\caption{Top three models for binary classification of data origin (Train or Test) indicating covariate shift}
\label{tab:covariate_train_test_table}
\begin{tabular}{p{2cm}p{2.4cm}p{2cm}p{2cm}p{2cm}}
\hline\noalign{\smallskip}
Model & Accuracy & F1 Score & Sensitivity & Specificity  \\
\noalign{\smallskip}\svhline\noalign{\smallskip}
XGBoost	& 0.99 & 0.99 & 0.97 & 1.00\\
RandomForest & 0.98 & 0.98 & 0.94 & 1.00\\
DecisonTrees & 0.97 & 0.98 & 0.96 & 0.98\\
\noalign{\smallskip}\hline\noalign{\smallskip}
\end{tabular}
\end{table}

\subsection{Feature Selection}
Conducting an initial performance analysis enables an evaluation of the suitability of individual features to detect instances of stress from physiological data. From this analysis, ECG demonstrates the highest stress detection accuracy using the hourly data denoted in Table \ref{tab:feature_selection_table}, which aligns with the frequent usage of ECG in psychological state detection approaches found in related literature \cite{Agrafioti,Akane_stress_mental_health, bakker_stress_gsr}. In addition, the inherent links between ECG and the ANS likely attribute the signal’s performance to indicating stress. Due to the prevalence of GSR in stress detection approaches \cite{bakker_stress_gsr, Santos_gsr}, higher classification accuracy was expected. The reduced performance compared with other approaches may stem from using different GSR features, recording methods or dataset normalisation operations.

Notably, the individual performance of deep ECG features achieves reduced classification accuracy and F1 scores. Likely as these features consist of reconstructed raw ECG signals, which provide a more granular perspective of ECG but may obscure the physiological changes associated with stress due to the number of data points and lack of a statistical summary highlighting the signals’ fluctuations.

A combined analysis utilising all features in a single input vector achieves moderate classification accuracy, lower than solely using ECG features, likely resulting from confusion caused by the Deep ECG features.

\subsection{Feature Importance}
\begin{figure}
    \centering
    \includegraphics[width=\textwidth]{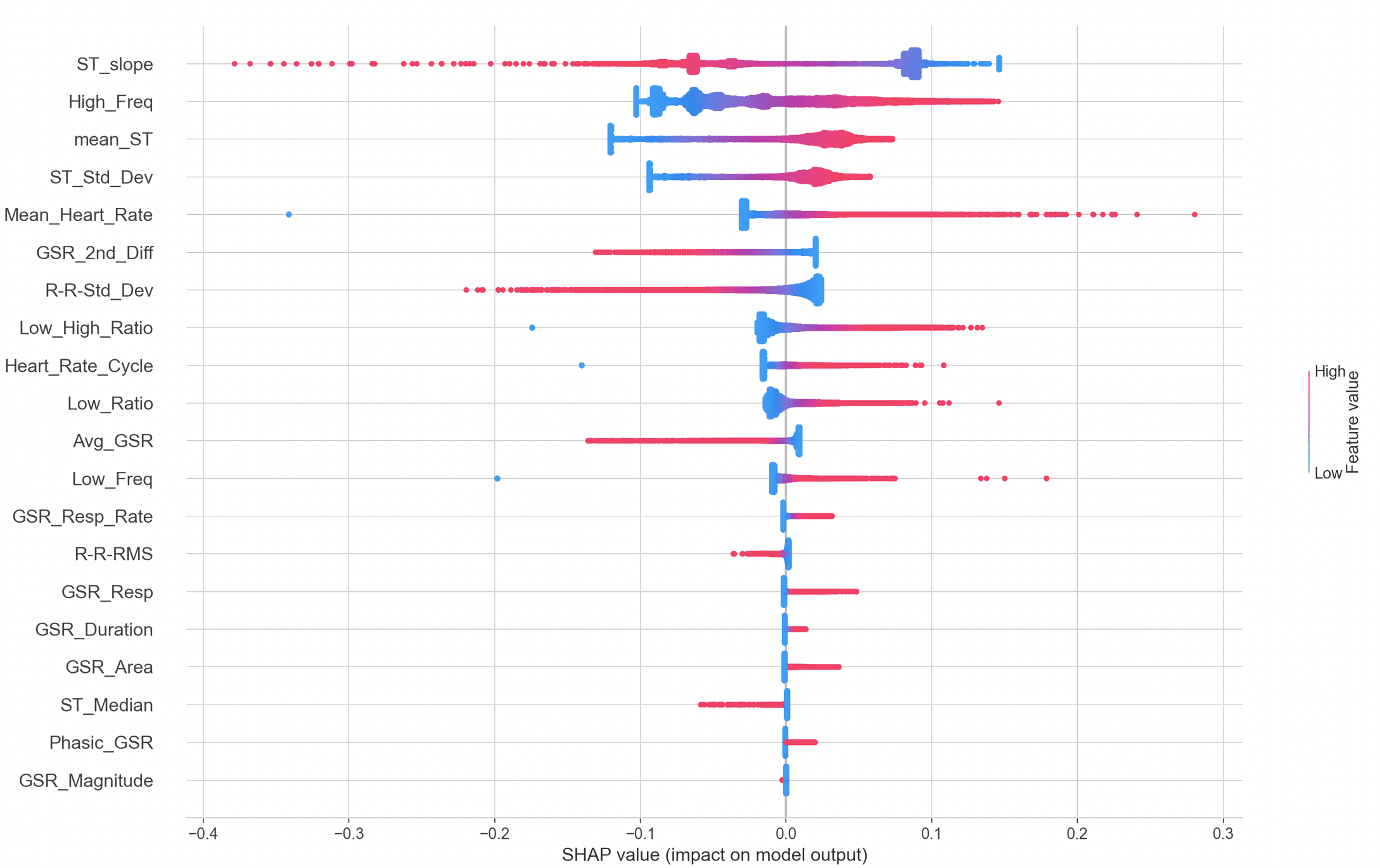}
    \caption{Summary of how the top features in a dataset impact the model’s output}
    \label{fig:shap_fi}
\end{figure}

The feature importance scores were calculated using SHAP based on the selected features that achieved good performance scores, namely handcrafted ECG and GSR features. The scores can be positive or negative based on whether the feature value positively or negatively impacted the positive classification (binary class = 1). A bee swarm plot in Figure \ref{fig:shap_fi} depicts the positive and negative importance scores assigned to the 20 handcrafted features in a logistic regression model. The plot shows the importance scores assigned across multiple instances in the data. The red-to-blue colour transition represents their high to low importance. 

It can be observed that there are many handcrafted features heavily impacting classification, the top five being the slope of the ST, high-frequency ECG signal, mean ST, the standard deviation of ST, and mean heart rate. Furthermore, the plot shows that the high slope of ST contributed towards negative stress classification. Contrastingly, high values of high frequency present in the ECG signal contributed toward positive classification as intuitively, heart rate might increase during stress. It was also observed that low mean ST and heart rate values contributed to negative classifications.

\subsection{Per-Minute Classification}
Per-minute classification reduces the data required to conduct momentary stress detection and provides a more granular analysis of the individual. Feature selection was also conducted on a per-minute basis. All features demonstrated an increased performance for stress detection, except for ECG, which exhibited a slightly reduced accuracy, Table \ref{tab:feature_selection_table}. 

Interestingly, the most performant features in per-minute classifications combine handcrafted ECG and GSR features. This is a result of the reduced dimensionality of the features, which places greater importance on the individual values comprising the ECG and GSR features, such as mean heart rate, heart rate cycle and skin temperature.

The increased performance can be attributed to per-minute instances providing isolated portions of data less likely to contain features indicative of both stress and neutral states. Furthermore, comparing each minute enables more significant distinctions between fluctuating data due to the granularity of the analysis. However, the reduced performance compared to hourly classifications can be explained by the greater importance placed on the individual features, emphasising the effect of missing values on classification.

\subsection{Feature Imputation}
Substantial portions of the SMILE dataset are missing features as denoted in Table \ref{tab:dataset_outline}. These missing values replicate a commonly occurring challenge in ambulatory assessments of physiological signals collected using wearable devices. Hence, warranting the adoption of two methods for feature imputation to approximate the missing values. 
The comparative performance of conducting feature imputation is demonstrated in Figure \ref{fig:results_feature_imputation}. Both methods result in a 1-2\% increase in accuracy for the combined ECG/GSR features, additionally sensitivity increases by 7\%, while a 3-4\% decrease in sensitivity occurs. While the performance on the training set shows a minor improvement, the Euclidean method contributed to the higher scoring achieved on the challenge data subset, with accuracy increasing by 3.75\% from 55.48\% to 59.23\%. Furthermore, the sustained performance using these methods indicates suitability to enable the replacement of missing data without negatively impacting the classifier, which provides a method for reducing the impact of sensor detachment in real ambulatory analysis.

\subsection{Generalisability}
A significant aspect of ML classification problems is achieving a generalisable and robust solution. Shuffling the training data enables the classifier to train on data randomly, originating from various individuals and reflecting stress and neutral instances. This method reduces bias and ultimately leads to a performant classifier, as demonstrated by the proposed approach in Figure \ref{fig:flowdiagram} and the ROC curve in Figure \ref{fig:results_roc_optimal_approach}. This proposed approach achieved high scores in all evaluation metrics: 90.77\% classification accuracy, 91.24 F1-Score, 90.42 Sensitivity and 91.08 Specificity. However, the same approach used to detect stress from the challenge dataset leads to 55.78\% classification accuracy. This substantial-performance disparity indicates significant variance between the training and testing data. 
Evidence of covariate shift is indicated by the high performance demonstrated for classifying between dataset splits denoted in Table \ref{tab:covariate_train_test_table}. This occurrence could arise in real-life applications for many reasons, such as physiological differences, noisy data, sensor variation, or nuanced physiological stress indicators. Therefore, emphasising the importance of comprehensive and balanced data collection for accurate psychophysiological analysis is critical.

\subsection{Challenge Approach}
The highest performing challenge approach achieves 59.23\% and is denoted in Figure \ref{fig:flowdiagram}. Achieved using the Euclidean feature imputation method on ECG and GSR features and an LDA model for per-minute classifications, with a second LDA classifier voting mechanism to provide hourly results.

\section{Conclusions}
An empirical evaluation aided the construction of a stress detection approach using physiological data collected from wearable technology in an ambulatory environment. The proposed approach leverages ECG and GSR features with dataset shuffling to ensure a random data distribution for training. The classifier utilised is ExtraTrees, which achieves high performance on a subset of SMILE training data across all evaluation metrics (90.77\% Accuracy, 91.24 F1-Score, 90.42 Sensitivity, 91.08 Specificity). Additionally, evaluated high-performing feature imputation methods are adopted to approximate missing data, which reduces the impact of issues such as uncontrolled sensor movement and possible detachment in ambulatory data-gathering settings.
Source code is available - https://github.com/ZacDair/EMBC\_Release.

\section{Future Work}
Due to the substantial performance disparity between SMILE training and testing data, future work will attempt to reduce covariate shift through multiple methods. An initial method relies on integrating a subset of testing data into the training dataset to evaluate the proposed approach’s capabilities for learning features of the test distribution. Subsequently, normalisation and re-weighting methods will be adopted to reduce the variance between the data.

%%%%%%%%%%%%%%%%%%%%%%%% referenc.tex %%%%%%%%%%%%%%%%%%%%%%%%%%%%%%
% sample references
% %
% Use this file as a template for your own input.
%
%%%%%%%%%%%%%%%%%%%%%%%% Springer-Verlag %%%%%%%%%%%%%%%%%%%%%%%%%%
%
% BibTeX users please use
\bibliographystyle{unsrt}
\bibliography{ref}

\end{document}